# Passive Classification of Source Printer using Text-line-level Geometric Distortion Signatures from Scanned Images of Printed Documents

Hardik Jain[+], Gaurav Gupta[+], Sharad Joshi, Nitin Khanna[*]

*Indian Institute of Technology Gandhinagar, India*

**Abstract**

In this digital era, one thing that still holds the convention is a printed archive. Printed documents find their use in many critical domains such as contract papers, legal tenders and proof of identity documents. As more advanced printing, scanning and image editing techniques are becoming available, forgeries on these legal tenders pose a serious threat. Ability to easily and reliably identify source printer of a printed document can help a lot in reducing this menace. During printing procedure, printer hardware introduces certain distortions in printed characters' locations and shapes which are invisible to naked eyes. These distortions are referred as geometric distortions, their profile (or signature) is generally unique for each printer and can be used for printer classification purpose. This paper proposes a set of features for characterizing text-line-level geometric distortions, referred as geometric distortion signatures and presents a novel system to use them for identification of the origin of a printed document. Detailed experiments performed on a set of thirteen printers demonstrate that the proposed system achieves state of the art performance and gives much higher accuracy under small training size constraint. For four training and six test pages of three different fonts, the proposed method gives 99% classification accuracy.

## 1. Introduction

With the rapid rise in advanced and sophisticated technologies, the printing industry has also witnessed much progress in the past decade. It has also coincided with the digitization of printed materials, but printed documents are still extensively used for many critical applications. The Confederation of European Paper Industries (CEPI), in its preliminary statistics report [1], estimates 410 million tons of paper production for the

---

year 2016 which has risen almost steadily over the past 15 years. Moreover, a research based on a global bottom-up model predicts that the global annual paper consumption is expected to rise to more than 600 million tons by the year 2030 [2]. Nonetheless, paper still serves an important purpose in various domains such as banks and legal tenders where it is hard to replace due to security concerns, time constraints, cost, and ease of use.

The total number of cheques that were cleared by the cheque clearing company of England and Wales in the year 2016 stood at more than 40 billion [3]. Also, the judicial system still runs on printed documents, and it would take a long time to replace the use of paper in developing countries. On the other hand, with new and refined technologies it is easier to forge documents, and hence there is a rising need to ascertain the authenticity of a printed document.

Researchers have shown a keen interest in developing automated systems to authenticate the genuineness of a printed document in question [4] (Section 2). While trying to make sure that a printed document is authentic, source identification of that document plays an important role [5, 6, 7, 8, 9, 10]. Source identification is also required in providing clues and leads in several criminal cases and in pinpointing the source of leakage of secure documents, counterfeiting, and forgery.

While printing a document, characters undergo some translational distortions about their initial locations [4]. Along with it, the character orientation is also affected by rotation distortion or skew up to $\pm 3\%$ and character size may also be compressed or expanded by around 2% [11]. This combination of character translation and rotation distortions is referred as geometric distortions (Section 3). These distortions are printer specific *i.e.* they vary across printers and depend on its make and model [6]. It is worth noting that translational and rotational distortions may also be introduced due to manual errors while printing and scanning. These distortions are usually uniform across the page and are not peculiar to a printer. The system proposed in this paper compensates such distortions during the pre-processing and pattern recognition stages. On the other hand, geometric distortions that can serve as printer's intrinsic signatures are location specific, but their change can be assumed to be fairly gradual across the page along with the printing direction.

This paper presents a novel system for printer classification using intrinsic signatures from character-level geometric distortions. Initially, an input scanned image undergoes a well-designed preprocessing which helps in minimizing distortions produced due to manual errors or errors from other sources which are not part of printer specific signature. In addition to the feature extraction stage, the proposed pre-processing stage is a major contribution of this paper. Character locations are then identified using optical character recognition followed by estimation of translational and scaling distortions of each printed character about its reference digital copy. These distortion values are used to estimate feature vectors corresponding to each printed line which are used for classification using support vector machine (SVM) classifier (Section 4). We have created a custom database of 13 printers for evaluation of the proposed method (Section 5.1). The database consists of English text documents printed from both laser and inkjet printers. Various experimental results show that the proposed method outperforms state-of-art techniques for printer classification (Section 5).

## 2. Related Works

Traditionally, source printer identification from hard copy documents is done using chemical analysis of ink [12, 13]. Over the past decade, digital image processing based systems are also evolving for source printer identification/classification using printed documents. Printer classification is carried out using unique print signatures namely extrinsic and intrinsic [14]. Extrinsic signature involves embedding an external signature in every printed document. Source printer classification and forgery detection can be done using this hidden information. In contrast to extrinsic signatures, intrinsic signatures are inherent to a printer due to different hardware and software involved in producing its output, are content independent and are unique for a printer [9]. These are invisible to naked eyes and are detected by scanning a printed document at high resolution and applying image analysis to extract features. Features proposed in this paper are based on intrinsic signatures, and previous works related to same are discussed in this section.

Some of the early works were based on analyzing the quality of printed documents to discriminate printing technologies [15]. Authors introduced several print quality metrics such as line width/raggedness, over spray, dot roundness/perimeter and number of satellite drops. Using these metrics, printing technologies were quantitatively analyzed in terms of statistically meaningful number of lines, halftone dots, and text features' parameters. These parameters were used to differentiate between digital and impact printing technologies. It is intuitively clear that for evaluation of these metrics, high-resolution document scanning is inescapable. Another approach involving high-resolution scan (2400 dots per inch (dpi)) was proposed in [16], [17], [18], [19] and [20]. This system explored banding artifacts as printer specific feature. The artifacts appear as alternating light-dark bands perpendicular to the printing direction and are due to quasi-periodic fluctuations in printing process direction. Banding frequencies for different printers were measured by printing mid-tone gray level patches created with line fill pattern. Banding was then modeled from actual document, to extract intrinsic parameters. However, the technique was suitable only for electrophotographic printers.

Identifying document source is an integral part of forgery detection, which was utilized in [21]. A feature vector for each character was formed by using four qualities: line edge roughness, area difference, correlation coefficient and texture. A character-wise decision was taken and used to detect forgeries in the test document.

Printed document degradation and noise was utilized by [22, 23, 24] for content independent character based printer identification. Document image degradation while printing was utilized as the main source of evidence by [22], from which four discriminative characteristics were identified: image noise and artifacts, character edge roughness, character edge contrast, and uniformity of printed character area. These characteristics were captured by statistical features related to noise, gradient, DCT, and multi-resolution wavelet analysis. A system using the difference in edge roughness for distinguishing laser and inkjet printed documents was proposed by [23]. Printed character edge roughness was estimated along the vertical edge by taking the standard deviation of pixel gray values. Distinctive noise introduced due to manufacturing imperfection was utilized by [24] for printer identification. These content independent techniques are sensitive to toner density and noise [9] and their performance also degrades if the document has partial content. With the development in technology, the noise spread of inkjet printers has reduced thereby reducing the applicability of these techniques in the modern scenario.

Geometric distortion features are more robust compared to content independent character level features, as they are independent of toner density. These intrinsic features are not affected by partial content [9]. Various techniques are available which had utilized geometric distortion but the scope of each method is restricted by various constraints. [25] analyzed geometric distortion in documents to classify different electro-photographic printers. A 2-D (dimensional) distortion displacement vector for each halftone dot position before and after printing is obtained and a collection of these vectors was used to form distortion signature for the printer. Correlation measure was used for signature similarity assessment. However, the technique primarily focuses on halftone images and has no mention of printed documents with text/characters only. Page distortion was modeled by [6] using projective transformation, based on the fact that geometric distortion can turn ideal parallel lines into intersecting ones. Character location from reference and printed document was used to find eight coefficients of projective transformation. Out of these eight coefficients, four coefficients which denote intrinsic features of a printer were used for laser printer classification. The technique lacks skew and/or offset compensation for manual mishandling while printing and scanning and is also limited to printed pages in specific languages with equispaced characters printed on them.

Geometric distortion features were further utilized by [4, 9] for forgery detection. Distortion mutation of geometric parameters was used in [4] for detection of forged characters in a printed document. After creation of reference document using OCR and rotation correction using Hough transform, translational distortion was estimated for the printed characters. Page-level geometric distortion features based on text lines were used by [9]. These included page text line slope (PTLS) and page text line interval (PTLI) as the horizontal and vertical features respectively. Concatenating slope of each best fitting text line in the document gives PTLS and concatenating vertical distance between adjacent lines gives PLTI. Based on these features, Euclidean distance was used as the similarity measure for classification. However, PTLS is sensitive to the presence of smaller lines, and PTLI requires the document to have a single paragraph.

Our proposed method caters to the shortcomings of above geometric distortion based techniques by introducing a novel set of features based on character-level distortions, preprocessing involving correction of manual mishandling, and creating a large dataset of English alphabets with multiple paragraphs in a page including small text lines as well. Proposed method performs row-level (classifying each printed line separately) classification, which is more versatile compared to existing page level accuracies reported on various datasets as this row-level (or line-level) classification scheme can also be extended to forgery identification in a printed document if one or more rows are forged. Amongst the existing geometric distortion based methods, projective transformation model based method presented in [6] is closest to the proposed system and hence it is used for comparison with the proposed system.

## 3. Geometric Distortion Features

This section provides a detailed description of geometric distortions in printed documents with sample illustrations from our database. Most of the commonly used electronic text documents in various domains such as legal paperwork, do not contain complicated formatting. These documents mostly contain text lines which are parallel to each other and have uniform spacing between them. Printing an electronic document involves a lot of miniature hardware equipment, and slight variations in different components of a printer might introduce distortions in a printed document. Most of the modern printers have inbuilt quality control mechanisms, and these variations in a printed document compared to its electronic version are small enough to be observed by naked eyes. As the primary objective of these printouts is to convey information to a human reader, printer manufacturers do not build mechanisms for correcting those printing defects which are not noticeable to naked eyes. However, when a printed document is scanned at high resolution, distortion in locations of printed characters as compared to their corresponding locations in the electronic version could be observed. The printed text lines undergo distortions and thus remain no more parallel and have non-uniform spacing between them. This distortion in printed documents relative to their electronic versions may be in the form of translation and rotation and is referred to as geometric distortion caused by the printer.

followed repeated who surprise. Great asked oh under on voice downs. Law together prospect kindness securing six. Learning why get hastened smallest cheerful. He 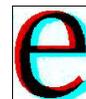

Figure 1: Overlaid characters of electronic document (in cyan) and printed document (in red), the overlapping region is shown in black color. Zoomed version of a character is shown on the right.

Figure 1 shows an example of character location distortions in a printed document with respect to its electronic version. Printed characters are shown in red color, its electronic version is denoted in cyan color, and the overlapping region is shown in black color. These two versions of a document are registered with respect to the location of the left-top corner of the bounding box corresponding to their first alphabet (for example 'f' in Figure 1). As we move along a document in either direction, slight deviations in printed character locations can be observed in the overlapped document. Such slight variation is not noticeable by naked eyes even if a printed document and its electronic version are kept next to each other. Figure 1 also shows a zoomed version of a single character 'e' to notice the geometric distortions that can be captured by a high-resolution scan of a printed document. The proposed system is based on utilizing such small geometric distortions in printed document which are unnoticeable by naked eyes but can be captured by automated digital image processing methods on a high-resolution scan of the printed document.

These geometric distortions are due to mechanical defects in printer hardware and vary along a page in both directions. During printing in an electrophotographic/laser printer, mechanical movements occur in polygon mirror and optical photoconductor drum. These components are responsible for reflecting the laser beam and transferring toner onto the paper, respectively. Any variation in their operation might introduce geometric distortions in printed documents [26]. Operating principle of inkjet printers significantly differs from those of laser printers. Inkjet printers involve mechanical back and forth movement of the printhead. Nozzle mounted on this printhead dispenses liquid ink onto a moving paper. Any misalignment in this mechanical movement would result in distortion in printed character's location and size [27].

Every printer uniquely affects the position and size of characters while printing [28]. Moreover, different brands of printers produce various types of geometric distortion signatures which are consistently retained for a long time [28]. These distortions in a printed document are quantized by estimating offsets in location and size of characters with respect to their electronic version. A shift in character location is referred to as translation distortion and change in the size of character as scaling distortion. These distortions form the backbone of our proposed features, and the efficacy of these features for printer classification has been illustrated in Section 5.

Geometric distortion is printer specific, but there might be some manual distortion introduced due to page mishandling such as manual mishandling resulting in incorrect paper feeding while printing and placing a paper in tilted position while scanning. Document pre-processing for translation and rotation correction is performed to eliminate these types of errors that do not characterize a printer (Section 4).

## 4. Proposed Method

Figure 2 shows an overview of the feature extraction step involved in the proposed system. Standard steps of training and testing a suitable classifier follow this feature extraction step. First of all, a printed document, as well as its corresponding electronic text (henceforth referred as reference document), undergo a strategic document preprocessing workflow. String matching between these two documents follows the preprocessing step. Finally, the locations of matched characters are used for estimating suitable features. Following subsections explain this procedure in detail.

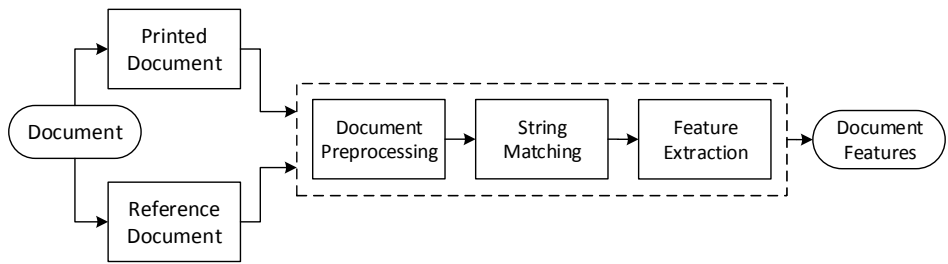

Figure 2: An overview of the proposed feature extraction scheme based on geometric distortions.

### 4.1. Document Preprocessing

Figure 3 shows different operations involved in the preprocessing of scanned and reference documents. For printed hard-copy document, it involves acquisition using scanner, followed by character identification and finally applying corrections. For the reference document, only translation correction is performed (highlighted in gray color in Figure 3). Reference document is a binary TIFF image denoted by $I_r(x, y)$. If the reference document is not readily available; it can be generated by either manually typing or by performing adjustments to the output of optical character recognition on a scanned version of the corresponding printed document.

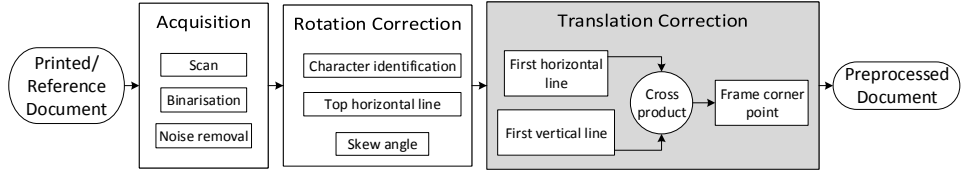

Figure 3: Block diagram of document preprocessing involved in the proposed system (reference document undergoes only the steps highlighted in gray color).

### 4.1.1. Acquisition

Printed document is scanned at 1200 dpi resolution and stored as $16-bit$ grayscale TIFF image, referred to as $I_s(x, y)$. This grayscale image $I_s(x, y)$ is converted to corresponding binary image $I_s^b(x, y)$ using Otsu's threshold [29]. Binary thresholding minimizes the effect of toner level variations across multiple pages. Because of hardware defects or print quality variations, printed documents might have ink spread, seen as impulse noise or satellite droplets. These noise components are selected using connected component analysis and components with less than $\gamma$ number of pixels are removed. The threshold $\gamma$ is selected based on the expected number of connected pixels in the correctly identified characters. This connected component based denoising is applied on $I_s^b(x, y)$ to obtain a denoised version $I_s^{bN}(x, y)$ of binarized scanned document.

### 4.1.2. Rotation Correction

Manual mishandling of paper while printing and scanning may incur an additional rotation apart from printer's characteristic distortion. This non-characteristic distortion needs to be removed as it will act as interfering noise for our goal of source printer identification. This paper proposes to eliminate this non-characteristic distortion by subjecting $I_s^{bN}(x, y)$ to rotation correction about the first text line. A rotation correction angle $(\theta_m)$ is estimated such that the first text line becomes parallel to scanned document's horizontal axis. Although, choosing the first line for mishandling correction would ignore any geometric distortion occurring in document area above the line. As there is no text above the first text line, neglecting that region does not affect the quality of final features used as signatures of the printer.

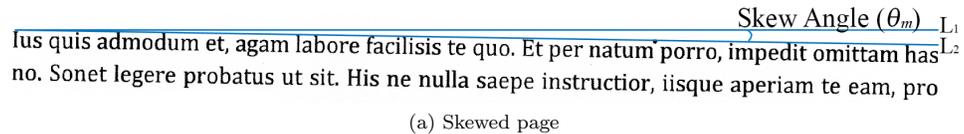

(a) Skewed page

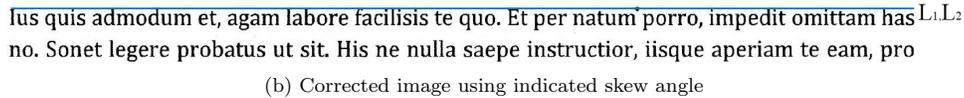

(b) Corrected image using indicated skew angle

Figure 4: Document skew correction (demonstrated on top two lines of a scanned document).

Figure 4a shows part of a scanned document before rotation correction and Figure 4b shows corresponding scanned document after rotation correction. The angle used for correcting the image is indicated in the top right corner of Figure 4a, as skew angle $\theta_m$. This rotation correction is carried out only for the scanned document as the reference document is free from rotation mishandling error. The skew angle $\theta_m$ is estimated using alphabet locations of the first text line. The identity of each character and its bounding box information is obtained from optical character recognition (OCR) using MATLAB's implementation of Tesseract OCR Engine [30]. This bounding box information contains rectangular character bounding box with base parallel to image horizontal axis. It includes information such as top left corner position, width, and height of the character (Figure 5a). This bounding box information is used to identify midpoint of the lower (in the vertical sense) horizontal boundary of a character (as marked in Figure 5b) in the document.

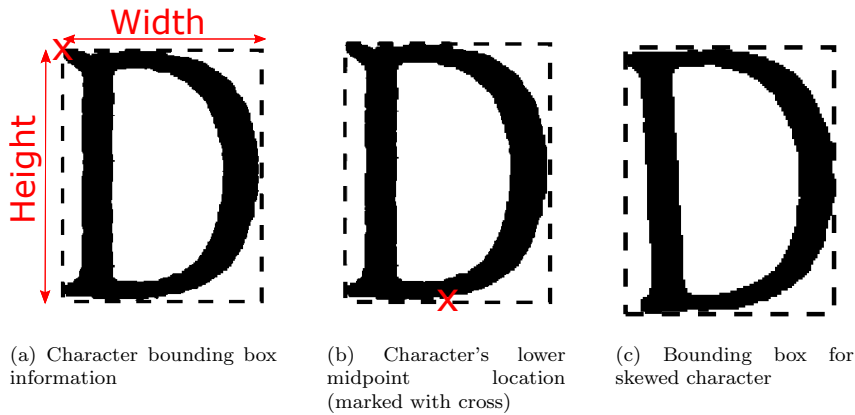

(a) Character bounding box information

(b) Character's lower midpoint location (marked with cross)

(c) Bounding box for skewed character

Figure 5: Character bounding box information

Rotation correction requires finding appropriate $\theta_m$ from text line character information. The lower boundary's midpoint of twenty-one English alphabets lie on the baseline, and only five ('g', 'j', 'p', 'q' and 'y') run below the baseline. To accurately determine $\theta_m$, only alphabets on the baseline are processed, and punctuation marks are also rejected. Let $L_1$ denote a set of lower boundary's midpoint locations of these alphabets in the first text line, $L_1 = \{(x_j, y_j) \,|\, j \in \{1, \ldots, N\}\}$. Here, $N$ is the number of alphabets (on the baseline) in the first text line. Then, the slope $m$ of fitted baseline can be estimated using least square fitting on $L_1$ locations (Equation 1) [31].

$$m = \frac{N \sum_{i=1}^{N} y_i x_i - \left(\sum_{i=1}^{N} x_i\right)\left(\sum_{i=1}^{N} y_i\right)}{N \sum_{i=1}^{N} x_i^2 - \left(\sum_{i=1}^{N} x_i\right)^2} \quad (1)$$

Rotation correction angle, $\theta_m = \arctan(m)$ is used for rotation correction of $I_s^{bN}(x,y)$ by $\theta_m$ rotation in anticlockwise direction. A similar approach has been utilized in [31], for skew correction of printed documents to correct distortions before document image analysis.

*4.1.3. Translation Correction*

On rotation corrected version of $I_s^{bN}(x,y)$, OCR is reapplied to find the character locations. From this information, first horizontal line $\vec{H}_1$ and first vertical line $\vec{V}_1$ in the homogeneous coordinate system are obtained. $\vec{H}_1$ is estimated from the $y$ coordinates of document's top text line alphabets, $\vec{H}_1 = [0 \quad 1 \quad -med(y_1, y_2, \ldots, y_h)]$. Similarly, $x$ coordinates from the left most alphabets of each text line are used to estimate $\vec{V}_1$, $\vec{V}_1 = [1 \quad 0 \quad -med(x_1, x_2, \ldots, x_v)]$. Here $med$ denotes median, $y_i \, \forall \, i \in \{1, \ldots, h\}$ and $x_j \, \forall \, j \in \{1, \ldots, v\}$ denote $y$ and $x$ coordinates of alphabet's bounding box lower midpoint lying on $\vec{H}_1$ and $\vec{V}_1$ respectively. Taking median instead of mean/average makes the proposed step robust to outliers which might occur because of possible indentation in starting lines of different paragraphs on the printed page. The number of alphabets on $\vec{H}_1$ and $\vec{V}_1$ lines is $h$ and $v$ respectively. The intersection of these two lines gives the frame corner point $c$, which in homogeneous coordinates is obtained by the cross product, $c = \vec{H}_1 \times \vec{V}_1$.

Rotation corrected $I_s^{bN}(x,y)$ is translated with respect to $c$ and the resultant, rotation-translation corrected binary image $I_s^{bNc}(x,y)$ is referred to as $I_s'(x,y)$. For consistency, similar translation correction step is carried out on $I_r(x,y)$ to give $I_r'(x,y)$ using its corresponding corner point. Both the preprocessed documents, $I_r'(x,y)$ and $I_s'(x,y)$ are simultaneously used for string matching.

*4.2. String Matching*

This paper proposes to estimate geometric distortion in printed documents using the correspondence between characters and their locations in printed and reference documents. One to one character matching is performed to calculate distortion across the document. Due to the limitations of OCR, sometimes all the characters are either not identified or not correctly classified. An exhaustive line-wise lowest common substring alphabet matching is performed [32], which ignores alphabets missing in either of the documents ($I_r'(x,y)$ and $I_s'(x,y)$). From these matched strings, alphabet box information $R_{ij}$ corresponding to $j^{th}$ alphabet on $i^{th}$ line (counted from top) of $I_r'(x,y)$ is extracted according to Equation 2.

$$\begin{aligned}
R = \{(x_{ij}, y_{ij}, w_{ij}, h_{ij}) | (i,j) \in \Psi\} \\
where \, \Psi = \{(1,1) \ldots, (1, J_1), \\
(2,1), \ldots, (2, J_2), \\
\vdots \\
(n,1), \ldots, (n, J_n)\}
\end{aligned} \quad (2)$$

Here, $R_{ij}$ is a four vector element for each alphabet with $(x_{ij}, y_{ij})$ being the coordinates of top-left corner of a character bounding box and $w_{ij}, h_{ij}$ being the width and height of the bounding box in $x$ and $y$ directions respectively (Figure 5a). The number of lines in a document is denoted by $n$ and the number of matched characters in $i^{th}$ line is denoted by $J_i$. In practice, the number of lines $n$ will vary from document to document and the number of characters in each line $J_i$ will vary from line to line. Total number of matched characters in a single document will be $\sum_{i=1}^{n} J_i$. Similarly, locations of corresponding matched characters in $I_s'(x,y)$ are obtained and referred to as $S(= \{(x_{ij}', y_{ij}', w_{ij}', h_{ij}')\})$. Thus, $R$ and $S$ are cardinal sets as the strings have already been matched.

### 4.3. Feature Extraction

Initial feature vector, $F_{R,S}$ is evaluated at reference image matched alphabet locations $\{(x_{ij}, y_{ij})\}$. Different values of this feature vector $F_{R,S}$ depend upon the sets $R$ and $S$ estimated earlier and each value is a four element vector for each matched alphabet pair (Equation 3). These vectors are translation and scaling distortions measured in $x$ and $y$ directions (Equation 4).

$$F_{R,S}(x_{ij}, y_{ij}) = (t_x(x_{ij}, y_{ij}), t_y(x_{ij}, y_{ij}), s_x(x_{ij}, y_{ij}), s_y(x_{ij}, y_{ij})) \tag{3}$$

$$\text{Translation distortion in } x \text{ direction}: t_x(x_{ij}, y_{ij}) = x_{ij} - x'_{ij}$$
$$\text{Translation distortion in } y \text{ direction}: t_y(x_{ij}, y_{ij}) = y_{ij} - y'_{ij}$$
$$\text{Scaling distortion in } x \text{ direction}: s_x(x_{ij}, y_{ij}) = \frac{w_{ij}}{w'_{ij}} \tag{4}$$
$$\text{Scaling distortion in } y \text{ direction}: s_y(x_{ij}, y_{ij}) = \frac{h_{ij}}{h'_{ij}}$$

The dimension of $F_{R,S}$ depends on the number of matched alphabets in a document, and it contains $\sum_{i=1}^{n} J_i$, four element members. These four parameters are evaluated at locations $\{(x_{ij}, y_{ij})\} \; \forall \; (i,j) \in \Psi$ (Equation 2). Since English alphabets are of different shapes and sizes and different lines of a document contain a varying number of alphabets. Therefore, these set of locations $\{(x_{ij}, y_{ij})\}$ are generally non-uniformly distributed on a document. An example set of alphabet locations $\{(x_{ij}, y_{ij})\}$ for a document is shown in Figure 6a, and when $t_x(x_{ij}, y_{ij})$ values are estimated at these locations, resulting feature is shown in Figure 6b.

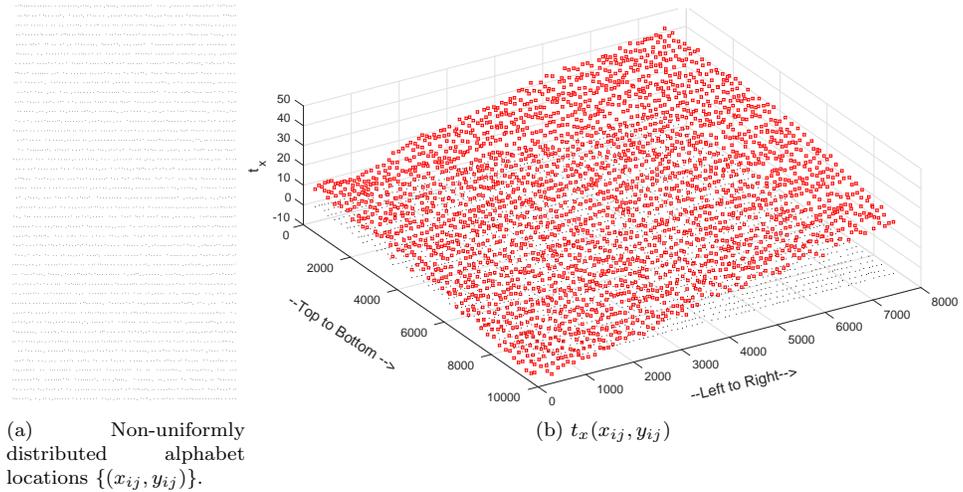

(a) Non-uniformly distributed alphabet locations $\{(x_{ij}, y_{ij})\}$.

(b) $t_x(x_{ij}, y_{ij})$

Figure 6: (a) Matched alphabet locations of a document and (b) $t_x$ values evaluated at this locations.

Considering $F_{R,S}$ as the final feature vector would make the feature dimension for different pages different. This inconsistent feature dimension would make the comparison of documents difficult as we will be comparing features evaluated on different sampling grids. So, $F_{R,S}$ which is a function of non-uniformly distributed locations $\{(x_{ij}, y_{ij})\}$, is reevaluated at uniformly distributed locations. For this four surfaces are fitted on 2-D scattered data using bilinear interpolation for $\{(t_x, t_y, s_x, s_y)\}_{ij}$. These surfaces are evaluated on uniform grid locations $\{(x_{rc}, y_{rc})\} \, \forall \, r \in \{1, \ldots, n_r\}, \, c \in \{1, \ldots, n_c\}$ (Figure 7a), which has $n_r$ lines and $n_c$ locations in each line. Figure 7b shows corresponding $t_x(x_{rc}, y_{rc})$.

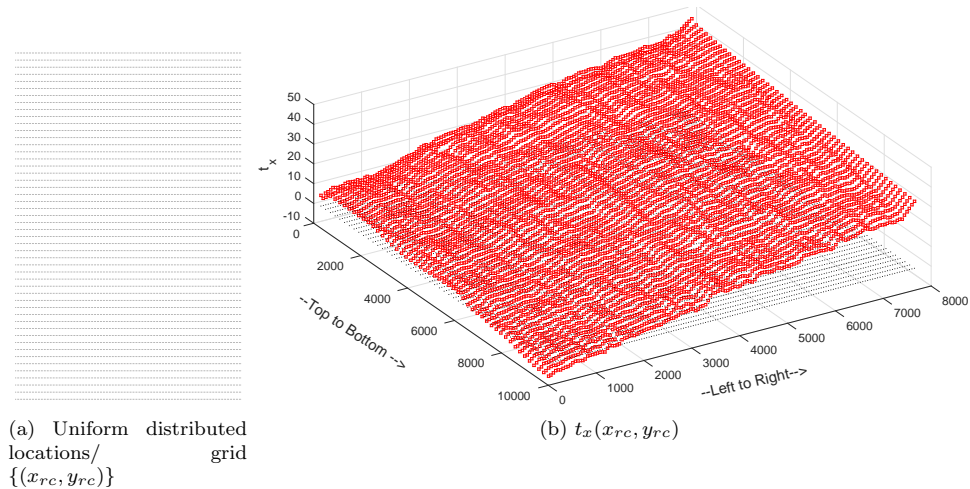

(a) Uniform distributed locations/ grid $\{(x_{rc}, y_{rc})\}$

(b) $t_x(x_{rc}, y_{rc})$

Figure 7: (a) Uniformly distributed locations on a document and (b) $t_x$ values evaluated at this dot locations

From a pair of printed and its corresponding reference document, obtained resampled feature vector, $F'_{R,S} = \{(t_x, t_y, s_x, s_y) | \, \forall \, r, c\}$, is referred as the independent feature vector of the printer. Since, translation distortions $t_x$ and $t_y$ depend on alphabets' locations on the document, whereas scaling distortions $s_x$ and $s_y$ depend on alphabets' size. Therefore, translation distortions $t_x$ and $t_y$ are expected to show more invariance towards variations in alphabets and noise as compared to scaling distortions $s_x$ and $s_y$. Alphabet's size estimated using bounding box given by OCR is much less robust to potential sources of noise such as skew in printed document which results in alphabet rotation. The computed bounding box of this rotated alphabet would be a rectangle with base parallel to the horizontal axis, which would be of a different size compared to the unskewed alphabet (Compare Figure 5b and 5c). Change in the dimension of new rectangular bounding box would largely depend on the alphabet. Hence $s_x$ and $s_y$ would have more impact of skew than its position $(x_{ij}, y_{ij})$ in the document. Thus, it is expected that $t_x(x_{rc}, y_{rc})$ and $t_y(x_{rc}, y_{rc})$ will have a more consistent behavior than $s_x(x_{rc}, y_{rc})$ and $s_y(x_{rc}, y_{rc})$ throughout the document and will have less intra-class variability. The performance of derived feature vectors resulting from the combinations of independent features is also evaluated to maximize the overall efficiency. These derived feature vectors are $t_{xy}$ and $t_{all}$ obtained by concatenating $t_x$, $t_y$ and $t_x$, $t_y$, $s_x$, $s_y$, respectively.

*4.4. Classification*

This paper uses one of the most widely used supervised classification technique in multimedia forensics, support vector machine (SVM) for printer classification. $C$-SVM implementation from LIBSVM library [33] with Gaussian kernel is used. The model trained using samples belonging to known classes when fed with unlabeled samples of same feature dimension, returns estimated probability of feature vector falling into the trained classes. The unlabeled sample is assigned to a class whose estimated probability is the highest. For each classification experiment, confusion matrix with true positive ($TP$) and true negative ($TN$) rates is evaluated from predicted and true class labels. Quantitatively the efficiency of features are compared in terms of average accuracy ($AUC$), $AUC = (TP + TN)/2$.

Instead of taking all the features from a complete document simultaneously, each row is separately considered. So a document would result in $n_r$ samples, each of dimension $n_c$. For the experiments reported in this paper, $n_r$ and $n_c$ are fixed as 50 and 150 respectively because all the pages in the dataset contain printed text of similar font size. These parameters are selected depending on the font size, their values can be chosen lower for a larger font size and larger for a smaller font size. Each independent feature vector fed to the classifier corresponds to a single text line of the uniform grid, lies in a $n_c$ dimensional feature space and is normalized to have zero mean. This way a document is classified into a particular printer based on the majority votes of $n_r$ labels. Section 5.3 shows the effectiveness of the proposed system when classified using row-wise features. Finally, for a printed document, accuracies of six classifiers (corresponding to $t_x$, $t_y$, $s_x$, $s_y$, $t_{xy}$, $t_{all}$) are evaluated.

## 5. Experimental Results

*5.1. Printer Database*

Earlier works in this field have used various printer databases consisting of 5 [6], 6 [25], 8 [9] and 10 [5, 8] different printers. The database used in [8] is the only publicly available database and has printed documents from 10 laser printers, all of them scanned at 600 dpi. Since the proposed system relies on high resolution scanned images to characterize minute geometric distortions and 1200 dpi resolution is available in most of the general purpose flatbed desktop scanners. Thus, we have created our own printer database owing to the lack of publicly available dataset meeting the requirements for our designed experiments. The printer database utilized in this paper consists of 13 printers of various types, brands, and models as listed in Table 1. In this table, first letter of a label denotes Printer type: Inkjet (I) or Laser (L) printer while the second letter denotes the brand: Epson (E), Brother (B), Canon (C), Hewlett-Packard (H), Konica Minolta (K), or Ricoh (R). It includes 12 Laser and an Inkjet printer. The size of the database is chosen to be bigger than those used in similar works in this research area, and printed documents are scanned at 1200 dpi resolution to better capture the geometric distortion features. We have printed 25 pages of three distinct fonts to demonstrate the efficiency of our features in different scenarios. Sets $P^{Ca15}$, $P^A$, and $P^{Co}$ denote fifteen pages of Cambria, five pages of Arial, and five pages of Comic Sans font from all the 13 printers. Different training and testing sets from these are formed and denoted by $P_{train}$ and $P_{test}$ respectively. A subset of randomly selected 5 pages from $P^{Ca15}$ is denoted by $P^{Ca5}$.

| Label | Model | Resolution (in dpi) |
|---|---|---|
| IE1 | L360 | 5760 × 1440 |
| LB1 | DCP7065DN | 2400 × 600 |
| LC1 | D520 | 1200 × 600 |
| LC2 | IR5000 | 2400 × 600 |
| LC3 | I6570 | 2400 × 600 |
| LC4 | LBP2900B | 2400 × 600 |
| LC5 | LBP5050 | 9600 × 600 |
| LC6 | MF4320D | 600 × 600 |
| LC7 | MF4820D | 600 × 600 |
| LH1 | Laserjet 1020 | 600 × 600 |
| LH2 | Laserjet M1005 | 600 × 600 |
| LK1 | Bizhub215 | 600 × 600 |
| LR1 | Aficio MP5002SP | 600 × 600 |

Table 1: List of printers used in this paper; first letter of label denotes printer type: Inkjet (I) or Laser (L) while the second letter denotes the brand: Epson (E), Brother (B), Canon (C), Hewlett Packard (H), Konica Minolta (K), or Ricoh (R).

5.2. Preliminary Analysis

In the proposed system, after uniform sampling each printed line/row is classified independently, henceforth referred as "row-level" classification (Section 4.4). "Page-level" accuracy (for classifying the whole printed page, considering all the printed lines/rows of the page together) is obtained by taking majority voting over all the decisions corresponding to different rows of the printed page. The first set of experiments are performed to compare the relative performance of the six parameters discussed in Section 4 and to obtain the optimal number of training pages for further classification tasks. Effect of training size on final classification accuracy is analyzed by choosing seven out of fifteen pages of $P^{Ca15}$ for testing. Then, six different classifiers are trained by randomly choosing $i$ ($i \in \{1, \ldots, 8\}$) pages out of remaining eight pages for training. All these classifiers are tested on the same testing set (randomly chosen seven pages of $P^{Ca15}$). Figure 8 shows row-level classification accuracy for different number of training pages of $P^{Ca15}$, while corresponding page-level accuracies are shown in Figure 9.

It is evident from Figure 8 and 9 that $t_x$ consistently gives the best performance out of the four independent features $t_x$, $t_y$, $s_x$, and $s_y$. Further, the combined features $t_{xy}$ and $t_{all}$ give very small improvement over $t_x$ and using $t_x$ requires much lesser computational resources then $t_{xy}$ and $t_{all}$. The performance of $t_x$ is very good even for a small number of training samples (3 or 4 pages per printer). This indicates that $t_x$ is the most suitable feature and also computationally faster than $t_{xy}$ and $t_{all}$ and the same is used for further classifications. Also, these figures indicate that there is not much increase in classification accuracy when the number of training pages is increased beyond 3. Therefore, for source printer identification from documents of the same font, three pages will be sufficient for training the model. This also indicates that our database size of five pages for other fonts will be sufficient to capture the variations in geometric distortion signatures across documents of the same font. Remaining same font experiments are performed by using three pages for training.

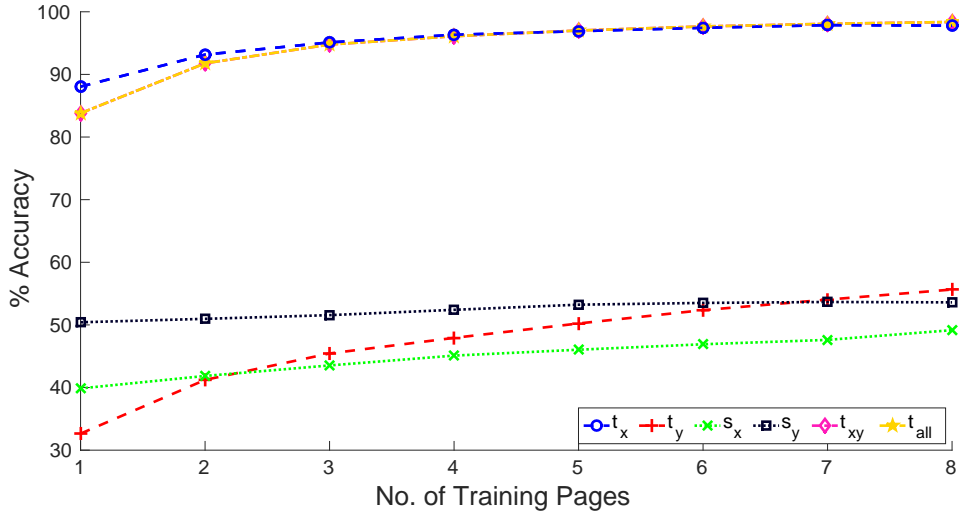

Figure 8: Row level classification accuracies for pages of $P^{Ca15}$

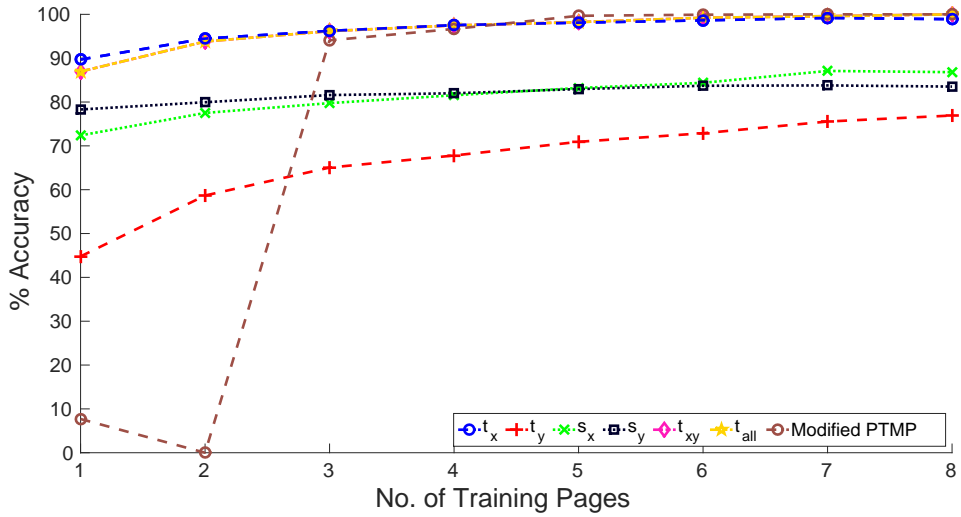

Figure 9: Page level classification accuracies for pages of $P^{Ca15}$.

Figure 9 also shows accuracy for printer classification using modified projective transformation model's parameter (PTMP) features proposed in [6]. Originally PTMP features were implemented on Chinese font only where all the printed characters across all the pages are of the same size, and each page contains an exactly same number of lines and number of characters per line. So, this paper uses a modified version of the method proposed in [6] because even for a fixed font type and size, the size of different English alphabets varies and a printed page will generally contain different paragraphs with some indentation (our database has these characteristics). Modifications include creating a similar uniform surface as that of proposed features and then calculating the PTMP features. As mentioned by authors, projective transformation model's parameter (PTMP) features yields 100% accuracy for classifying ten printers by using six pages from each printer for training and another six pages for testing. Our proposed modification to PTMP (henceforth referred as "Modified PTMP") also achieves close to 100% accuracy for six training pages from each of the thirteen printers and another seven pages for testing. However, the proposed features give close to 100% accuracy (both row-level and page-level) using only four pages per printer for training. This makes the proposed system more practical than the system based on modified PTMP features because printing and scanning more pages, especially when the number of printers is large, will be a very resource-consuming task and might not always be feasible.

### 5.3. Classification Results

Similar to the classification results for $P^{Ca15}$, two different classifiers are separately trained for $t_x$ features obtained from printed documents containing text in two other fonts $P^A$ and $P^{Co}$. Training-testing pairs have been created according to Equation 5, where $P_i^A$ and $P_i^{Co}$ denote $i$ pages from $P^A$ and $P^{Co}$ respectively for training. $P_{2t}^A$ and $P_{2t}^A$ denote the two testing pages of $P^A$ and $P^{Co}$ respectively. These two test pages are first randomly selected out of the five pages and kept fixed across all training sizes; then the training pages are chosen from the remaining non-overlapping set of three pages. Row-level and page-level accuracies corresponding to the training-testing sets are mentioned in Table 2.

$$\{(P_{train}, P_{test})\} = \{(P_i^A, P_{2t}^A), (P_i^{Co}, P_{2t}^{Co}) \mid i \in \{1, 2, 3\}\} \\ \text{for } P_i^A \cap P_{2t}^A = \emptyset,\ P_i^{Co} \cap P_{2t}^{Co} = \emptyset \tag{5}$$

| $P_{train}/P_{test} \downarrow$ | | Row level | Page level | Row level | Page level | Row level | Page level |
|---|---|---|---|---|---|---|---|
| No. of training pages $\rightarrow$ | | 1 | | 2 | | 3 | |
| Proposed Features | $P^A$ | 93.33 | 94.87 | 95.23 | 96.15 | 95.38 | 96.15 |
| | $P^{Co}$ | 90.72 | 92.31 | 94.62 | 97.44 | 98.08 | 100 |
| Modified PTMP Features | $P^A$ | - | 7.69 | - | 0 | - | 76.92 |
| | $P^{Co}$ | - | 7.69 | - | 0 | - | 69.23 |

Table 2: Classification accuracies (in %, $t_x$ parameter) for training-testing on different number of pages of $P^A$ and $P^{Co}$ for proposed and modified PTMP features. Same training-testing pages were used for each instance of proposed as well as modified PTMP features.

From results of Figure 8, 9 and Table 2, it can be concluded that for large training size (at least six pages from each printer) the proposed features perform as good as the modified PTMP features for printer classification on same font. While for small training data (such as three pages from each printer), the proposed features give much better classification accuracies than the modified PTMP features. The proposed features give good classification accuracies even with a single training page.

In practice, a large training set of a particular font might not always be available, and classification has to be carried out with limited training data. In order to evaluate the performance of proposed features for such scenario, training-testing pairs from $P^{ALL} = \{P^A \cup P^{Ca5} \cup P^{Co}\}$ are created. Training-testing pairs are formed according to Equation 6. The testing set $P_{6t}^{ALL}$ contains an equal number of randomly chosen pages of each font (two from each font). This training set is fixed for testing all the classifiers in this experiment while the training set $P_i^{ALL}$ is created by randomly choosing $i$ out of remaining nine pages and repeating this random selection over $^9C_i$ iterations. Row-level classification accuracies over $^9C_i$ iterations for $i$ number of pages is illustrated by error plot in Figure 10. Corresponding page-level accuracy statistics are depicted in Table 3, along with their comparison with modified PTMP features.

$$\{P_{train}, P_{test}\} = \{(P_i^{ALL}, P_{6t}^{ALL})| \ i \ \in \ \{1,\ldots,9\}\} \\ \text{for } P_i^{ALL} \cap P_{6t}^{ALL} = \emptyset \tag{6}$$

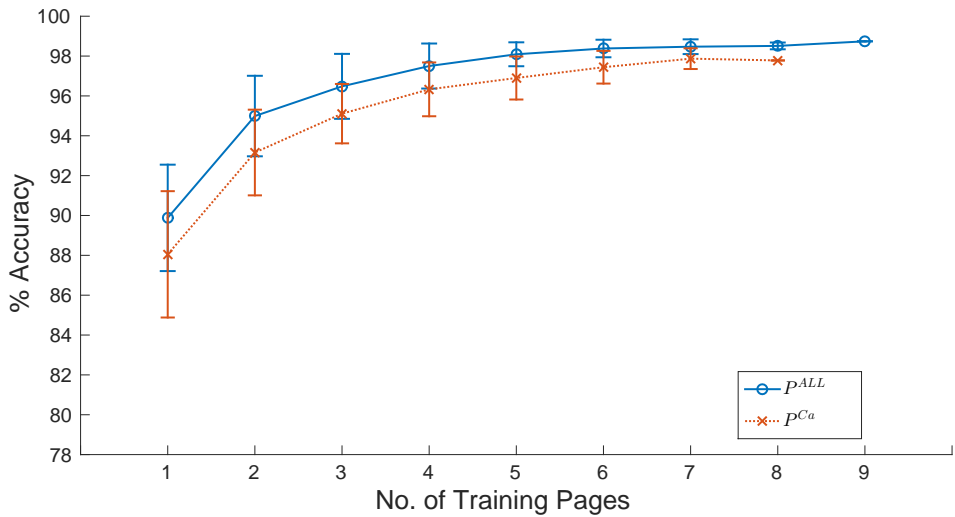

Figure 10: Error plot for row-level accuracy of proposed features for different number of training pages of $P^{ALL}$ and $P^{Ca}$.

Figure 10 indicates that classification accuracies for the proposed features increases with increase in the number of training pages and features give good performance even at a very small number of training pages. Results further improve for page level classification (Table 3). Table 3 shows that classification accuracies corresponding to the modified PTMP features also increases with increase in the number of training pages, but they are unable to give good performance for small training size. For example, for three training pages, the classification accuracy of the proposed features is more than 4% higher than the corresponding accuracy of the modified PTMP features, and the standard deviation is also less for the proposed features.

|  | No. of Training Pages → | 1 | 2 | 3 | 4 | 5 | 6 | 7 | 8 | 9 |
|---|---|---|---|---|---|---|---|---|---|---|
| Proposed Features | Mean ($\mu$) | 92.45 | 96.69 | 98.13 | 99.05 | 99.73 | 99.91 | 99.96 | 100 | 100 |
|  | Standard Deviation ($\sigma$) | 3.04 | 2.38 | 2.03 | 1.52 | 0.72 | 0.40 | 0.21 | 0 | 0 |
| Modified PTMP Features | Mean ($\mu$) | 7.69 | 0 | 93.97 | 97.23 | 99.03 | 99.56 | 99.79 | 100 | 100 |
|  | Standard Deviation ($\sigma$) | 0 | 0 | 5.43 | 3.87 | 1.72 | 0.87 | 0.48 | 0 | 0 |

Table 3: Page level accuracy statistics (in %) for different number of training pages of $P^{ALL}$ for proposed and modified PTMP features.

Printer-wise classification accuracy for a particular scenario is illustrated in Figure 11. Classification accuracy bar graphs at four training pages for all 13 printers are shown. In each case, first two bars corresponds to row-level and page-level accuracies of the proposed method respectively, and the third bar shows the page-level identification accuracy of modified PTMP features. Except for printers $LC1$ and $LC2$, for all other eleven printers, proposed features perform better than modified PTMP features and their ability to perform row-level classification makes them much more useful as they can also be extended for forgery detection in printed document when few rows of a printed document are forged. Hence, the experimental results presented in this paper show that the novel step of converting non-uniform grid to uniform grid extends existing PTMP features, removes their restriction to a single language and makes them applicable to all other languages as well. Over a large database, the proposed system outperforms existing systems for geometric distortion based printer classification and has potential to be extended for text-line level forgery detection as well.

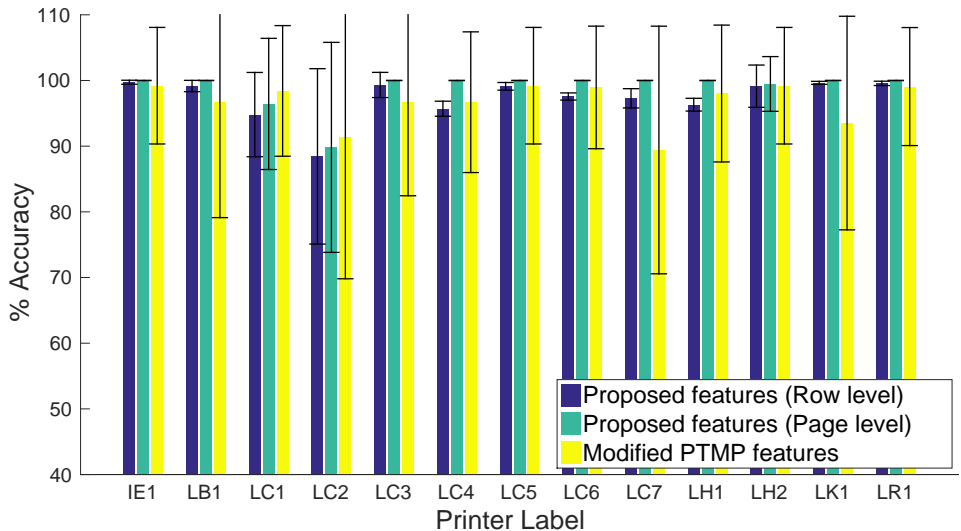

Figure 11: Average printer classification accuracy with error bars for four training pages of $P^{ALL}$. Same training-testing pages were used for each instance of proposed as well as modified PTMP features.

## 6. Conclusion

Geometric distortions in different characters of printed documents intrinsically exist on every page printed by a printer. Although the quality control mechanisms of modern printers make them unnoticeable by naked eyes, they can still be detected on high-resolution scans, such as 1200 dpi scans of documents printed at 300 dpi (commonly used printing resolution for printing text documents). Thus, if precisely modeled these geometric distortions can be used to classify printed documents and even detect forgeries in them. This paper proposed a system for estimating geometric distortions at row-level and using them for source printer classification from printed documents, needing only small amount of training data. Manual error/mishandling, if any, while printing or scanning, is addressed in the pre-processing stage. In addition to the pre-processing stage, the novelty of this work lies in projecting the geometric distortion parameters of a printed document from a non-uniformly spaced grid of characters to a uniform grid. This projection makes the algorithm content independent and will enable an easy extension of this technique to any language. Proposed method builds a printer model from a novel set of features estimated from its printed documents using SVM. Features of printed test document are compared with all these models using SVM and classification is performed based on prediction probability. An extensive experimental analysis shows that a printer model can be built and trained using any font from any language. Classification results show improved page-wise printer identification compared to existing techniques. From the experimental results on 13 printers, it is observed that the proposed method with four training pages gives 99.05% classification accuracy on six test pages, this accuracy increases further as the number of training pages increases.

Unlike existing literature, this paper proposes a row-wise detection. So, a document with text lines printed from different printers can be identified as forged. Hence, future work will include extending the proposed method for printed document forgery detection, at a computationally lower cost. Estimation of geometric distortion requires reference document, which if not available, can be generated from printed document either by typing or using OCR.

## Acknowledgment

This research was supported by the Board of Research in Nuclear Sciences (BRNS), Department of Atomic Energy (DAE), Government of India (Sanction Number DAE-BRNS-ATC-34/14/45/2014-E and Visvesvaraya PhD Scheme, Ministry of the Electronics & Information Technology (MeitY), Government of India. Any opinions, findings, and conclusions or recommendations expressed in this material are those of the author(s) and do not necessarily reflect the views of the funding agencies. Address all correspondence to Nitin Khanna, nitinkhanna@iitgn.ac.in.